*Pixel Multiplexing for high-speed multi-resolution fluorescence imaging.*

Gil Bub[1], Mathias Tecza[2], Michiel Helmes[1], Peter Lee[1] & Peter Kohl[1]

[1]Department of Physiology Anatomy and Genetics, [2]Department of Physics, University of Oxford, Oxford, UK. *contact: gil.bub@dpag.ox.ac.uk*

**We introduce a imaging modality that works by transiently masking image-subregions during a single exposure of a CCD frame. By offsetting subregion exposure time, temporal information is embedded within each stored frame, allowing simultaneous acquisition of a full high spatial resolution image and a high-speed image sequence without increasing bandwidth. The technique is demonstrated by imaging calcium transients in heart cells at 250 Hz with a 10 Hz megapixel camera.**

Conventional cameras capture images at a fixed spatial and temporal resolution. Events that occur faster than the camera frame integration time (T*i*) appear blurred (if moving) or as an average light intensity (when still). We propose a pixel multiplexing (PM) paradigm for imaging that allows simultaneous high-speed and high-resolution image capture[1]. The method adds a high-resolution shutter array that can open and close at speeds that create pixel expose times T*e* shorter than T*i,* so that exposure of detector subregions is independently controllable. A high-speed image sequence can be embedded in a single stored frame using the following scheme (Fig.1):

1) The image detector integrates and reads out a frame in the standard fashion, every T*i* seconds.
2) A shutter array produces a pattern of open/closed exposures during T*i* so that spatially dispersed pixel clusters (each consisting of one or more pixels) are exposed at the same time for $Te = Ti \times n^{-1}$.
3) Pixels exposed at the same time can be combined to form a single lower-resolution image. A series of *n* lower resolution sequentially exposed subframes can be extracted from the stored frame to produce a high-speed image sequence (*n*-times faster) from the original high-resolution frame.

PM embeds high-speed temporal information in an image while increasing neither memory requirements nor the intrinsic frame rate of the camera. Static regions of an image appear unchanged, while high-speed events are multiplexed into regions of the image that would normally appear blurred.

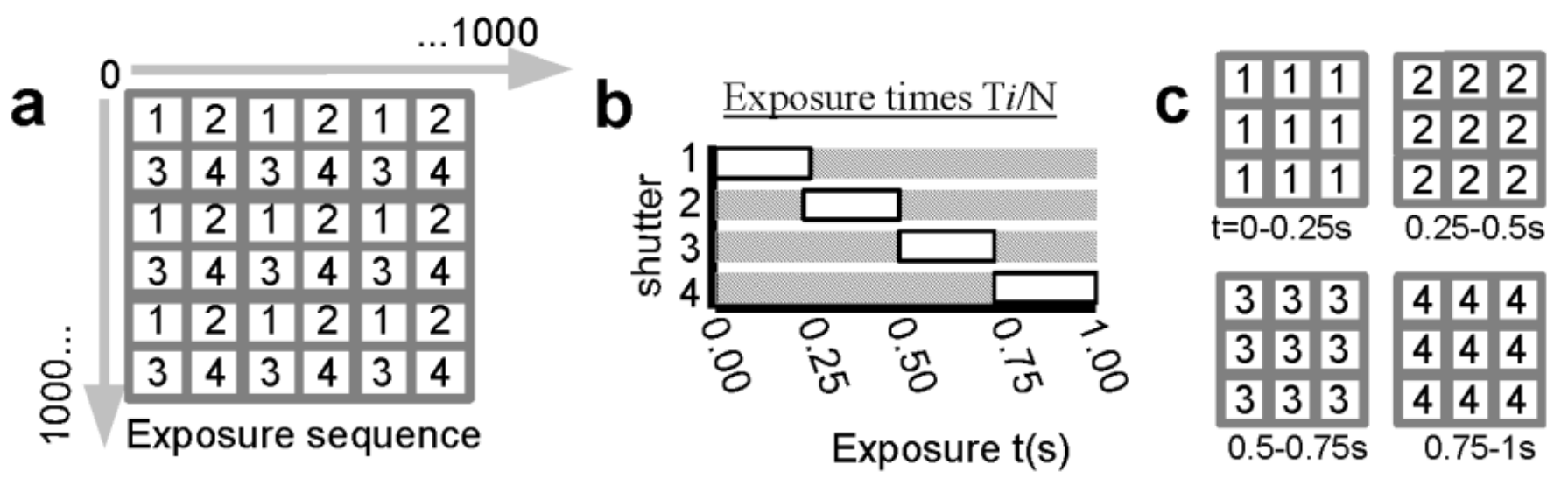

*Figure 1: Example of the proposed method with square (2x2) exposure groups. a) A megapixel CCD detector with a maximum frame rate of 1 frame per second is placed directly behind an in-register megapixel electronic shutter. b) The shutter produces a pattern of exposures so that 2x2 pixel exposure groups integrate light in sequence as shown. The pattern is repeated for each of the 250,000 exposure groups over the entire megapixel field of view. c) A frame representing the first 0.25s of the CCD exposure is constructed from (all) pixels#1 from each exposure group, a frame representing 0.25-5s is constructed from pixels#2 , and so on, to give 4 sequential 500x500 pixel frames collected within one second (4 fps). The original full resolution megapixel image is simultaneously obtained.*

There are several ways of implementing subregion exposure control. A principal design requirement is that spatially distinct subregions of the imaging chip are sequentially exposed faster than the frame rate of the camera would allow. Control can be done on the illumination side, using patterned light illumination techniques such as those applied to super-resolution[2] and confocal[3] imaging. Alternatively, the image can be blocked on the detector side, as done in metrology[4], dynamic filtering[5], or single pixel cameras[6]. Finally, PM could be incorporated into new CMOS chip designs, simply by offsetting pixel sample and digitization times. Here we present a proof of principle design, using a DMD micromirror array to transiently expose subregions of a CCD.

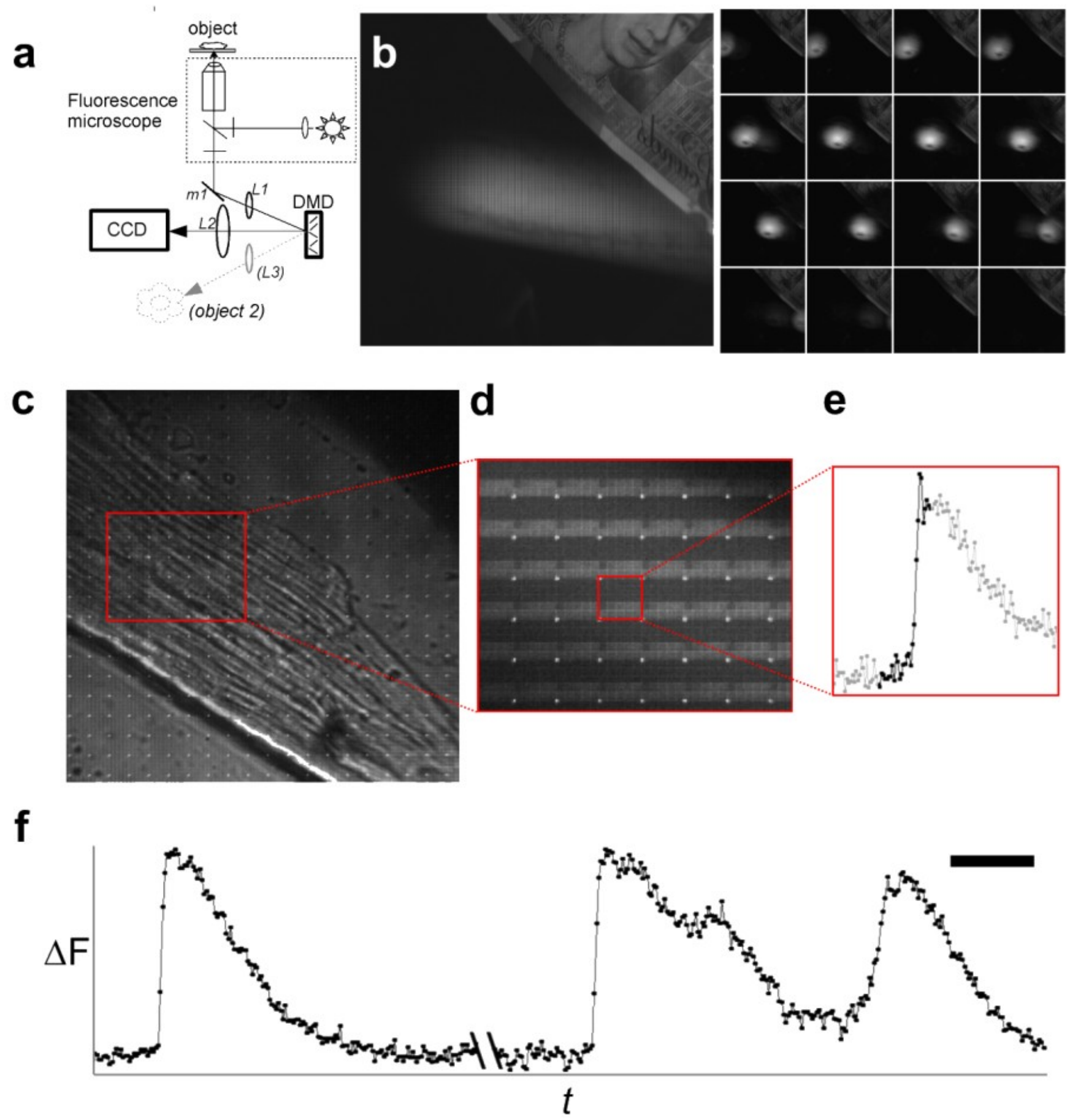

*Figure 2. Working prototypes: a) A schematic of the prototype, which can capture microscopic (black lines, via L1) or macroscopic (gray lines, via L3) images. b) High resolution (1000x1000 pixels) and 16 high speed embedded (250x250 pixels each) frames of a ball moving rapidly across a scene, captured via L3. Static high resolution details (currency in background) remain visible, while high speed (200 fps) image sequence is resolvable. c) Cardiac cell loaded with calcium sensitive dye, imaged with a megapixel camera at Ti=100ms (10 Hz). The DMD is programmed to expose a 5x5 pattern of 25 element pixel clusters for Te=4ms. Bright points on the image are mirrors toggled to the 'always on' position, for alignment purposes. d) Close up of cell regions during an action potential, showing intensity increases mapped to detector location. Every exposure group (such as the one outlined in red) measures fluorescence intensity from a 25x25 pixel area. e) Intensity vs. time plot for one of the exposure groups. Dark points correspond to pixel values in (d). f) Average intensity values for all pixels in (d) clearly resolve the calcium transient shape. A normal beat is followed at a later time by one with an early and delayed after contraction. Differences in upstroke velocity are clearly resolvable. Scale bar 150ms.*

We have developed a working prototype (Fig.2a) based on a machine vision camera (Prosilica GC1380H) and a XGA+ DMD mirror array running at 16kHz (Texas Instruments). The camera is focused on the mirror array so that there is a 1:1 correspondence between pixels and mirrors. The imaged scene is projected so that individual DMD mirrors act as shutters, allowing selective exposure of camera pixels. Camera and mirror array are positioned to allow imaging from the side port of a microscope (Nikon Diaphot, 40x oil immersion lens) or for macroscopic imaging using suitable projection lensing (e.g. lens L3 in Fig.2a). A demonstration of PM in action is shown in Fig.2b. A high-resolution image of a macroscopic scene was imaged while the DMD array cycled through a sequence of 16 mirror patterns every 80 ms ($T_e$=5ms), for a final frame rate of 200 fps. After image decoding, the blur in the high-resolution image is resolved as a ball rapidly moving across the frame.

The method is particularly useful for functional imaging studies, such as conducted in cardiac[7] and neuronal[8] cell-networks. Here, tissue motion isn't measured, rather cells undergo rapid ion concentration and membrane voltage changes which can be mapped to light intensity using fluorescent probes. Since high-resolution detectors suffer noise and bandwidth limitations at high frame rates, and since image intensification technology imposes dynamic range limits[9,10], researchers typically rely on high-speed, high dynamic range, low resolution detectors, ranging from 16x16 photodiode arrays[7], to specialized low resolution CCD[8] and CMOS[11]cameras for data collection. Additional high spatial resolution cameras are often used to allow interrelation of structure (cell morphology and anatomy) and function.

We conducted a proof-of-principle functional optical mapping study in heart tissue. Isolated rat cardiac myocytes were loaded with the calcium sensitive dye Rhod-2 (Invitrogen) and imaged at 40x using an inverted microscope. High-resolution images (Fig.2c/d) were captured while simultaneously measuring whole cell calcium transients at high speed (Fig.2e/f). The effective frame rate of calcium transient capture is more than an order of magnitude greater than that of the camera itself (here 250fps vs. 10fps), and significantly exceeds the cameras' maximum region of interest (ROI) and on-chip binning frame rates. Further, the fluorescent transient S/N is high; in our hands it is equivalent to that obtained from specialized high-speed low-resolution devices, such as photodiode arrays, dedicated high-speed CCDs, or photomultiplier tubes.

PM has several benefits compared to conventional approaches. First, by simultaneously imaging two resolutions with one detector, we remove alignment and light loss issues associated with splitting the light path to multiple detectors, and reduce the memory/bandwidth requirements as both resolutions are embedded in one image. The data throughput requirements are therefore the same as that of the single high-resolution image sequence, which allows use of slow scan cameras, with their known cost, signal to noise and dynamic range advantages[10,12], for high-speed imaging tasks. The system is also highly flexible: our present implementation allows for non-rectangular ROI shapes (pixel groups in the shape of cells, for example), as well as use different frame rates for different areas of the CCD, which is not possible using conventional detectors. Furthermore, using suitable wavelength switching, it is possible to monitor different functional characteristics of a sample simultaneously, and/or relate those to underlying structure.

We thank Philip Cobden and Richard Vaughan-Jones for isolated cells.